%% file: cdfgambc.tex
\documentclass[prl,twocolumn,showpacs,superscriptaddress,floatfix,preprintnumbers]{revtex4}

\usepackage{graphicx}
\usepackage{dcolumn}

\begin{document}





\title{Measurement of the cross section for direct-photon production in association with a heavy quark in $p\bar{p}$ collisions at $\sqrt{s}$ = 1.96 TeV}
\input{Author_List}
\preprint{FERMILAB-PUB-13-082-E}




\begin{abstract}
We report on a measurement of the cross section for direct-photon production in association with a heavy quark using the full data set of $\sqrt{s}=1.96$ TeV proton-antiproton collisions corresponding to 9.1 fb$^{-1}$ of integrated luminosity collected by the CDF II detector at the Fermilab Tevatron. The measurements are performed as a function of the photon transverse momentum, covering photon transverse momentum between 30 and 300 GeV, photon rapidities $|y^{\gamma}|<1.0$, heavy-quark-jet transverse momentum $p_{T}^{jet}>20$ GeV, and jet rapidities $|y^{jet}|<1.5$. The results are compared with several theoretical predictions.
\end{abstract}


\maketitle

The study of direct-photon ($\gamma$) production in association with a heavy quark $Q$ ($b$ or $c$) in hadronic collisions provides valuable information on the probability distributions of partons inside the initial-state hadrons. At photon transverse energies $E_{T}^{\gamma}$ \cite{co} smaller than 100 GeV,  such events are produced predominantly by the Compton scattering process $gQ\rightarrow\gamma Q$, while at higher energies the dominant process is quark-antiquark annihilation with gluon ($g$) splitting to heavy quarks $q\bar{q}\rightarrow\gamma g\rightarrow\gamma Q\bar{Q}$ \cite{Stavreva:2009vi}. It is conventional to assume that the charm ($c$) and bottom ($b$) quarks in the proton arise only from gluon splitting. However, there are other models that allow the existence of intrinsic heavy quarks in the proton \cite{Brodsky:1981se}. A cross-section measurement of $\gamma+Q+X$ ($X$ can be any final-state particle) production provides information on the heavy-quark and gluon parton distribution functions (PDFs) and on the rate of final-state gluon splitting to heavy quarks.

The Collider Detector at Fermilab (CDF) Collaboration studied the process $p\bar{p}\rightarrow\gamma+b+X$ at $\sqrt{s}$ = 1.96 TeV, for photons in the range $20<E_{T}^{\gamma}<70$ GeV \cite{Aaltonen:2009wc}. The measured cross section agreed well with a prediction based on a perturbative quantum chromodynamics (pQCD) expansion \cite{Stavreva:2009vi} at next-to-leading order (NLO) in the strong coupling constant $\alpha_{s}$. The D0 Collaboration measured the cross section for photons in association with heavy-flavor jets using data collected at $\sqrt{s}$ = 1.96 TeV, covering the range $30<E_{T}^{\gamma}<300$ GeV \cite{Abazov:2012ea}. The results disagreed with the NLO pQCD prediction for both bottom jets and charm jets in the region $E_{T}^{\gamma} \gtrsim 70$ GeV. 

In this Letter, we present the updated CDF measurements of the cross sections of photon with heavy-flavor jets, using the full data set from 9.1 fb$^{-1}$ of integrated luminosity collected by the CDF II detector, exploring $E_{T}^{\gamma}$ up to 300 GeV with improved techniques. The CDF II detector \cite{Acosta:2004hw} has a cylindrical geometry with approximate forward-backward and azimuthal symmetry \cite{co}. It contains a tracking system consisting of silicon microstrip detectors and a cylindrical open-cell drift chamber immersed in a 1.4 T magnetic field parallel to the beam axis. The silicon subsystem is used for reconstructing charged-particle trajectories (tracks) and heavy-flavor-decay vertices displaced from the primary interaction point. Electromagnetic (EM) and hadronic calorimeters surrounding the tracking system with pointing-tower geometry are used
to measure photon energies. At a depth approximately corresponding to the maximum development of the EM shower, the EM calorimeters contain fine-grained detectors (central electromagnetic strip chambers) that measure the shower profile. Drift chambers and scintillators located outside the calorimeters identify muons. 

The data are collected using a three-level online event-filtering system (trigger) that selects events with at least one energy cluster consistent with a photon in the final state. The trigger is approximately 100\% efficient for signal events in the explored kinematic region. The offline event selection requires the primary vertex $z$ position to be within 60 cm of the center of the detector. Each event is required to have at least one photon candidate that  has pseudorapidity \cite{co} in the fiducial region of the central calorimeter (approximately $|\eta| <$ 1.04). The transverse energy of the photon is corrected to account for non-uniformities in the calorimeter response, and calibrated using electrons from reconstructed $Z$ boson decays. Photon candidates are required to have $E_{T}^{\gamma} >$ 30 GeV and to satisfy pre-selection requirements on calorimeter and tracking isolation and the ratio of the energy measured in the hadronic calorimeter to the EM energy, as described in Ref. \cite{Aaltonen:2009in}. An artificial neural network (ANN) is constructed from calorimeter- and strip-chambers-shape information and isolation variables to further reduce background \cite{Aaltonen:2011rj}. The photon candidates are required to pass a suitable threshold on the ANN output (0.75) for optimal signal-to-background discrimination.

At least one jet must be present in each event. Jets are reconstructed using the JETCLU algorithm \cite{Abe:1991ui} with a cone radius $R = \sqrt{(\Delta\phi)^{2}+(\Delta\eta)^{2}} = 0.4$ in the azimuthal angle $\phi$ and pseudorapidity $\eta$ space \cite{co}. We select jets that have $E_{T} >$ 20 GeV and $|\eta| <$ 1.5. At least one jet is required to be classified as a heavy-flavor jet using a secondary-vertex tagger \cite{Acosta:2004hw}. This tagging algorithm exploits the long lifetime of hadrons containing $b$ or $c$ quarks, and is based on the reconstruction of a displaced, or secondary, vertex using the reconstructed tracks. 
If multiple-tagged jets are present in the explored kinematic region, the one with the highest $E_{T}$ is selected. The selected jet is required to be reconstructed in a volume outside an $\eta-\phi$ cone of $R = 0.4$ surrounding the photon candidate. 

\begin{figure}[htp]
\begin{center}
\includegraphics[width=1.0\linewidth]{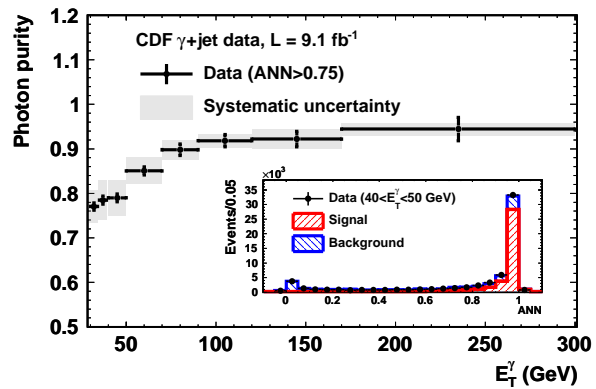}
\end{center}
\caption{Photon purity as a function of $E_{T}^{\gamma}$ for events restricted to the ANN signal region. The fit to the ANN distribution for photon candidates passing pre-selection requirements and with $E_{T}^{\gamma}$ between 40 and 50 GeV is shown in the inset.}
\label{fig:photopur}
\end{figure}

After all the selection requirements, 214 336 events remain in the data sample. Two main background sources contribute to these events: jets misidentified as photons (false photons) and light-flavor jets mimicking heavy-flavor jets. To estimate the rate of false photons, the photon ANN distribution in data is fitted to a linear combination of templates for photons and jets, obtained from a simulated inclusive photon sample using {\sc sherpa} \cite{Gleisberg:2008ta} and a dijet sample using {\sc pythia} \cite{Sjostrand:2006za}, after applying all the photon selection criteria except the requirement on the ANN output. 
The photon and jet templates are validated using the $Z^{0}\rightarrow e^{+}e^{-}$ and dijet data samples, respectively. 
A fit is performed in each $E_{T}^{\gamma}$ interval, yielding prompt photon fractions (purities) between 77\% and 94\% in the ANN signal range. The resulting photon purities and one example fit are shown in Fig.~\ref{fig:photopur}. The systematic uncertainties on the photon purities are estimated by varying the input variables to the ANN within their uncertainties. The dominant uncertainty on the shape of the ANN originates from the modeling of calorimeter isolation energy. The overall uncertainty is estimated to decrease from 6\% at $E_{T}^{\gamma} = 30$ GeV to 2\% for $E_{T}^{\gamma} > 70$ GeV.

\begin{figure}[htp]
\begin{center}
\includegraphics[width=1.0\linewidth]{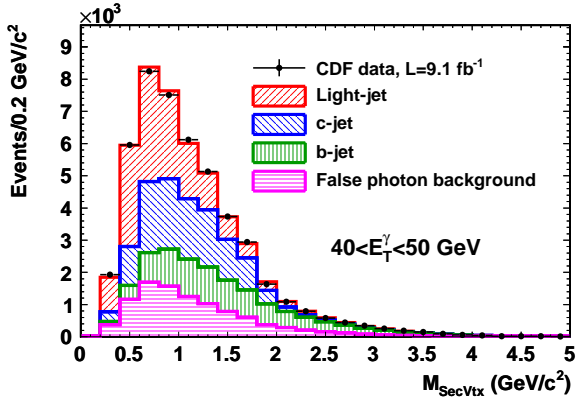}
\end{center}
\caption{Distribution of the secondary-vertex mass of tagged jets after applying the full selection, for photon candidates with $40 < E_{T}^{\gamma} < 50$ GeV. The points are data, and the stacked, shaded histograms represent the estimated contributions from the fit of the $b$-, $c$-, and light-quark jets and false photon background.}
\label{fig:msv}
\end{figure}

Backgrounds to heavy-flavor jets arise from light-flavor jets where random combinations of tracks mimic a displaced vertex. The fractions of $b$- and $c$-jets are determined by fitting the invariant mass ($M_{\rm SecVtx}$) of the system of charged particles, assumed to be pions, originating at the secondary vertex, using the templates for $b$-, $c$-, and light-quark jets constructed with {\sc pythia} \cite{Sjostrand:2006za}. The contribution to the $M_{\rm SecVtx}$ distribution from events with a false photon is modeled using dijet data, where one jet is required to deposit most of its energy in the EM calorimeter to mimic a photon and the other jet is required to pass all the heavy-flavor-jet selection. The loose photon requirement selects predominantly false photons. This background component is then constrained to the number of false photons from the ANN fits. After subtracting the contribution from events with false photons, 22\% to 37\% of the observed tagged jets are $b$-quark jets, and 16\% to 24\% of the observed tagged jets are $c$-quark jets for $E_{T}^{\gamma}$ between 30 and 300 GeV. The systematic uncertainties range from 15\% to 30\% and are dominated by the uncertainties in the simulated $M_{\rm SecVtx}$ template shapes originating from the uncertainty in the modeling of tracking-system efficiency. Figure~\ref{fig:msv} shows the result of the fit for $E_{T}^{\gamma}$ between 40 and 50 GeV, as an example.


The differential cross section as a function of $E_{T}^{\gamma}$ is defined as $d\sigma^{b(c)}/dE_{T}^{\gamma} = Nf_{\gamma}f_{b(c)}/(\Delta E_{T}^{\gamma}\epsilon_{\rm trig}\epsilon_{\rm UF}\mathcal{L})$, where $N$ is the number of data events in a given $E_{T}^{\gamma}$ bin after applying the full selection, $f_{\gamma}$ is the photon purity, $f_{b(c)}$ is the $b$-jet ($c$-jet) fraction in events with true photons, $\Delta E_{T}^{\gamma}$ is the $E_{T}^{\gamma}$ bin size, $\epsilon_{\mbox{trig}}$ is the trigger efficiency, and $\mathcal{L}$ is the integrated luminosity. The bin-by-bin unfolding factor $\epsilon_{\rm UF}$ combines corrections for acceptance, efficiencies of the photon selection and tagging algorithm and resolution effects to infer the results at the hadron level, using prompt-photon events simulated with {\sc sherpa} \cite{Gleisberg:2008ta}. The numerator of the unfolding factor is obtained by applying the same requirements to the {\sc sherpa}-simulated events as the ones applied to data. The denominator of the unfolding factor is obtained by applying the same kinematic and isolation selection on the generated quantities. Unfolding factors obtained with {\sc pythia} \cite{Sjostrand:2006za} are used to evaluate the systematic uncertainties. The photon efficiency is calibrated by comparing the selection efficiencies for $Z^{0}\rightarrow e^{+} e^{-}$ events in data and in simulation. The tagging efficiency is calibrated with data enriched with heavy-flavor jets. The unfolding factors range from 18\% to 27\% for $\gamma+b+X$ events and from 4\% to 8\% for $\gamma+c+X$ events. The systematic uncertainties are estimated to be approximately 10\% and are dominated by the uncertainties in the photon-energy scale and the tagging efficiency.

\begin{figure*}[htp]
\includegraphics[width=7.5cm]{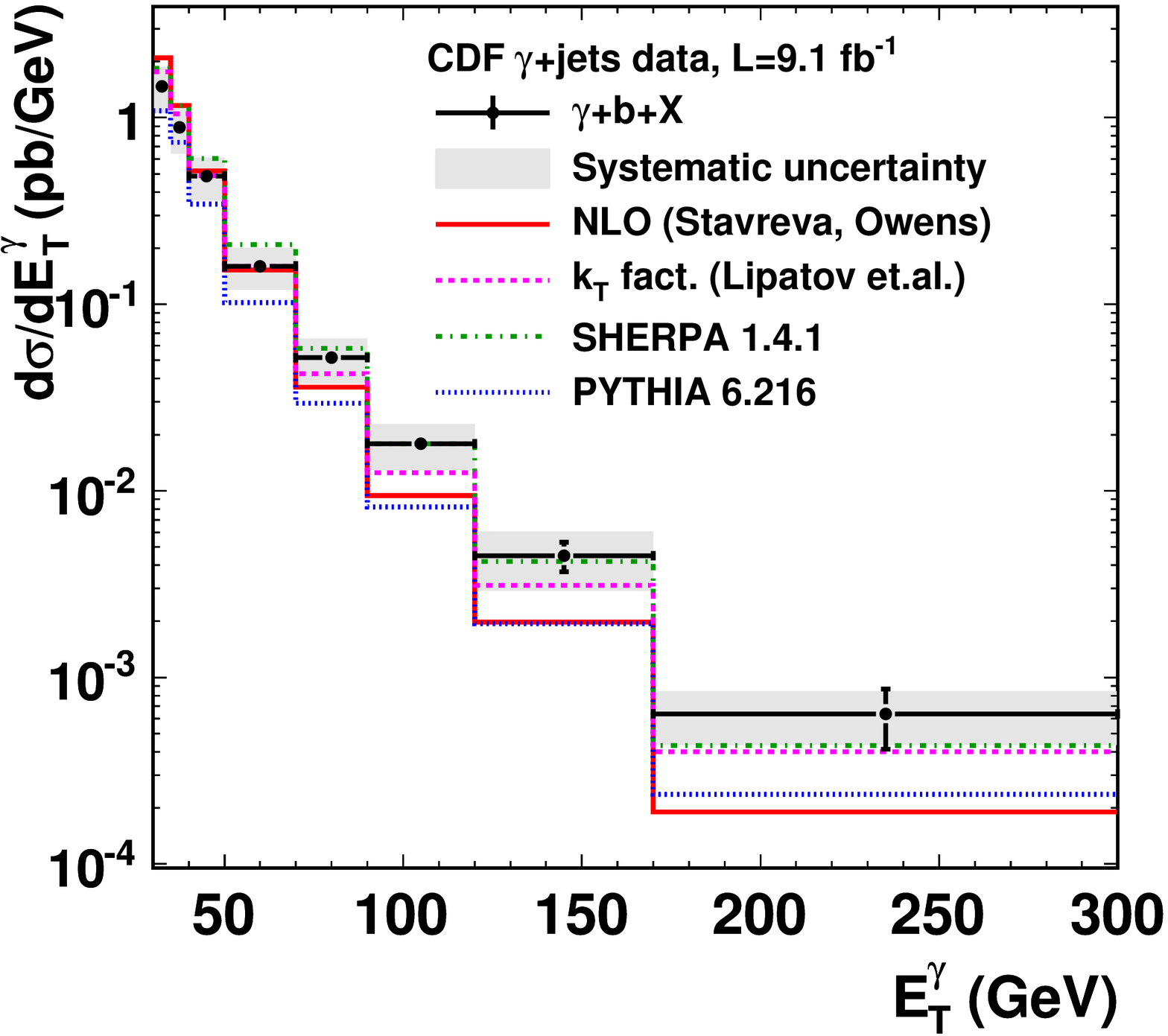}\hspace*{0cm}
\includegraphics[width=7.5cm]{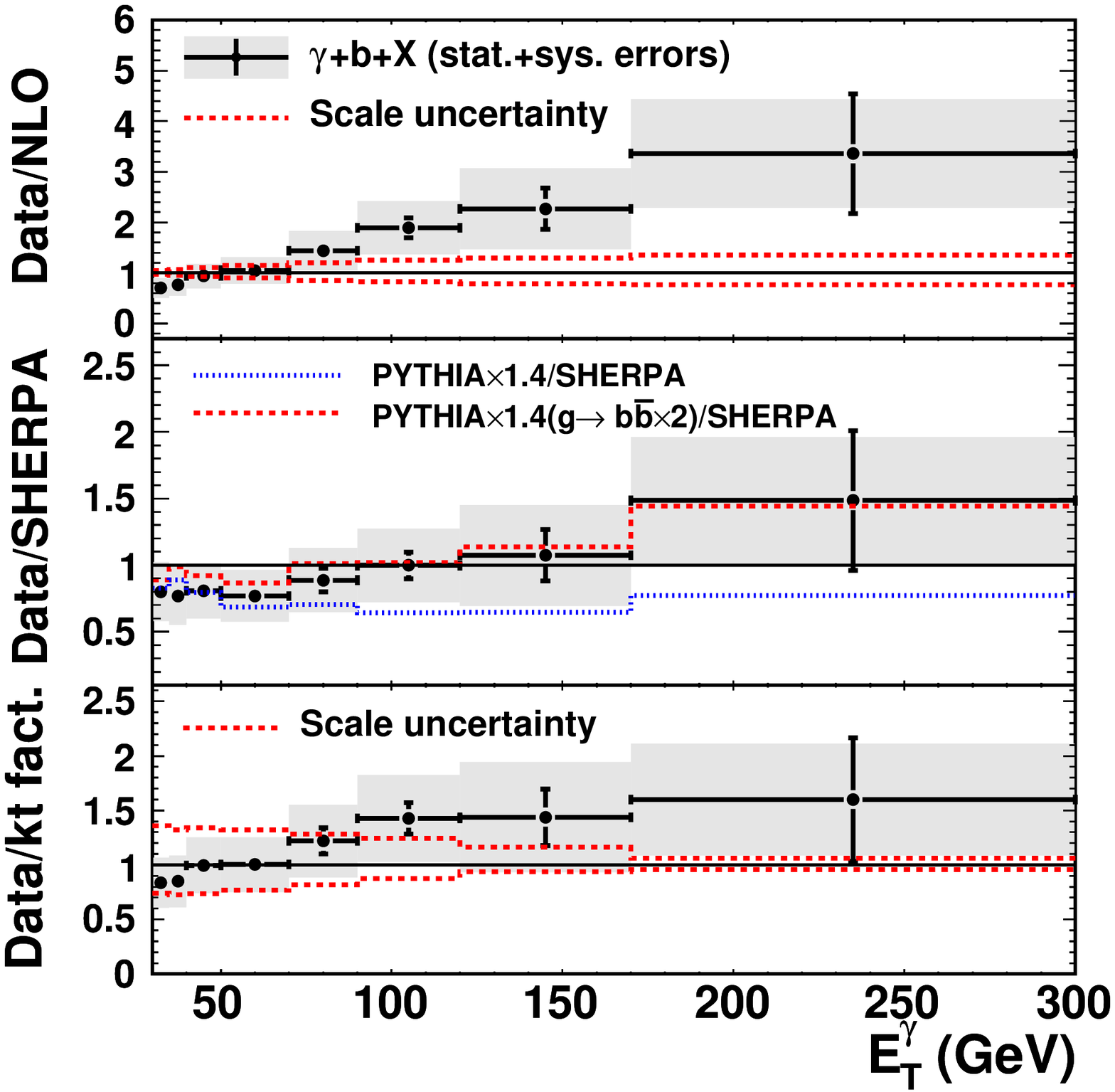}
\vspace*{0cm}
\includegraphics[width=7.5cm]{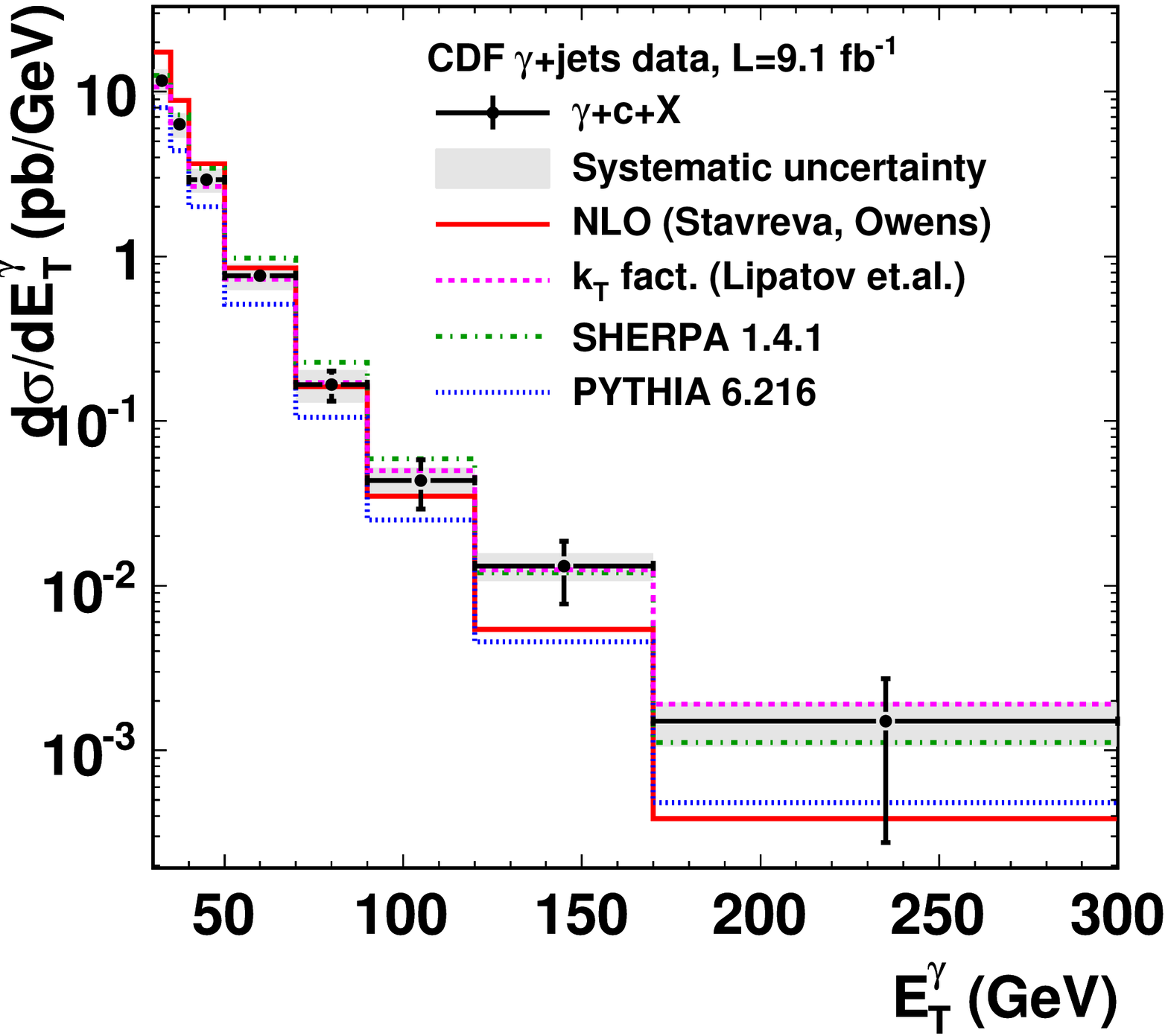}\hspace*{0.35cm}
\includegraphics[width=7.5cm]{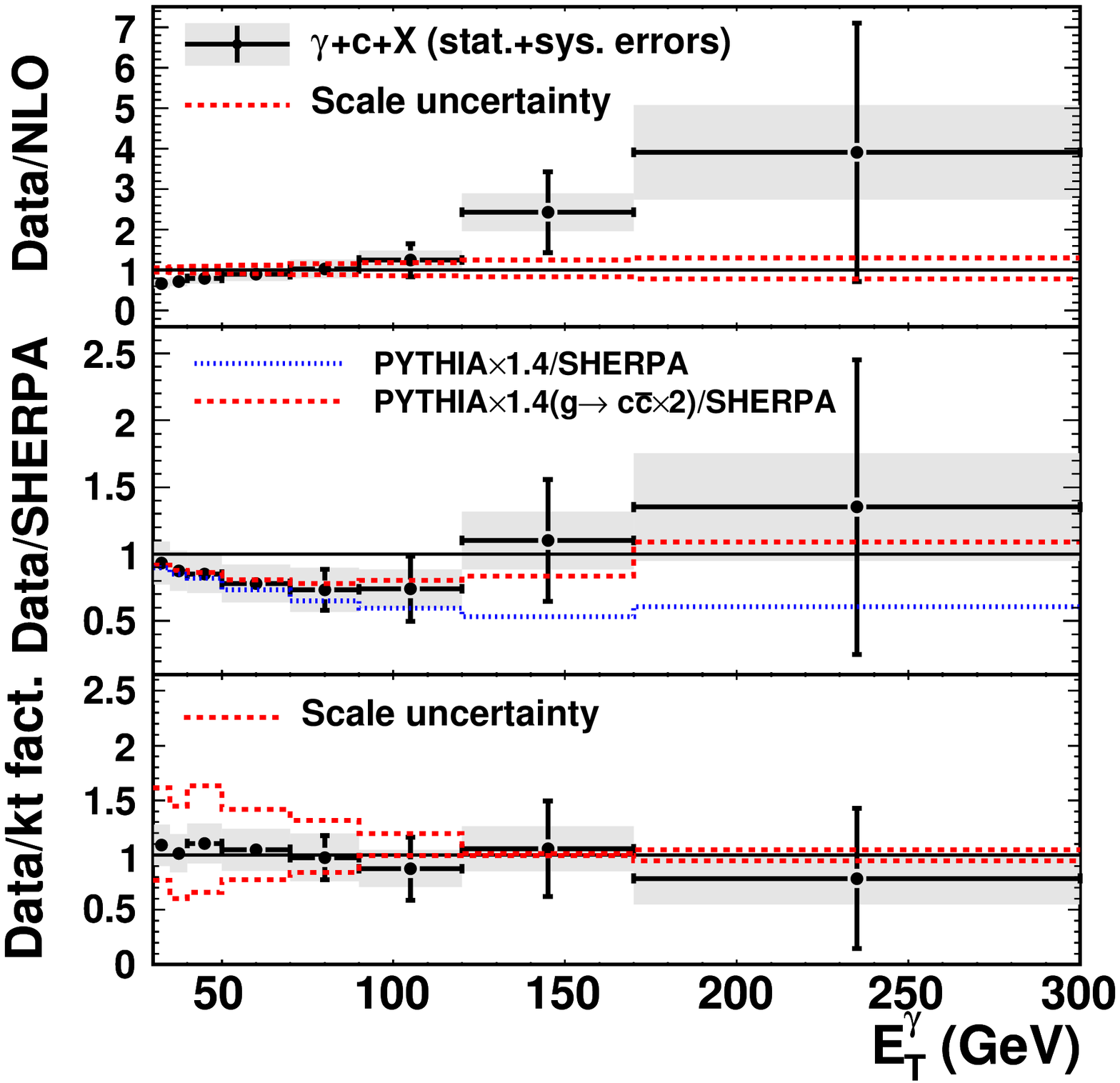}
\caption{The measured differential cross sections compared with theoretical predictions. The left panels show the
absolute comparisons and the right panels show the ratios
of the data over the theoretical predictions. The {\sc pythia} predictions are scaled by 1.4 in the ratio distributions. 
The comparisons are shown for $\gamma+b+X$ (top) and $\gamma+c+X$ (bottom) processes. The shaded area
around the data points indicates the total systematic uncertainty of the
measurement. The scale uncertainties are shown for the NLO and the $k_{T}$-factorization predictions.\label{fig:results}}
\end{figure*}

The measured differential cross sections for $\gamma+b+X$ and $\gamma+c+X$ productions and four theoretical predictions are listed in Table~\ref{tab:results} and shown in Fig.~\ref{fig:results}. The sources of systematic uncertainty on the integrated cross sections are summarized in Table~\ref{tab:systematics}.

\begin{table*}[htp]
\begin{center}
\caption{The $\gamma+b+X$ and $\gamma+c+X$ cross sections in intervals of $E_{T}^{\gamma}$ together with statistical and systematic uncertainties. Four theoretical predictions are shown. The scale uncertainties are shown for the NLO and the $k_{T}$-factorization predictions. The {\sc sherpa} and {\sc pythia} predictions have large scale uncertainties, which are not shown in the table. The last column shows non-perturbative corrections applied to the NLO and the $k_{T}$-factorization parton-level predictions.}
\begin{tabular}{c c  c  c  c c c}
\hline\hline
$E_{T}^{\gamma}$ bins & $d\sigma/dE_{T}^{\gamma}$ & NLO \cite{Stavreva:2009vi} & $k_{T}$ fact. \cite{Lipatov:2012rg} & {\sc sherpa} \cite{Gleisberg:2008ta}& {\sc pythia} \cite{Sjostrand:2006za}& Corr.\\
(GeV) & (pb/GeV) & (pb/GeV) & (pb/GeV) & (pb/GeV) & (pb/GeV) &\\
\hline
\multicolumn{7}{c}{$\gamma+b+X$} \\
\hline
30-35&1.47$\pm$0.07$\pm$0.41&$2.09\pm0.10$&$1.76\pm0.64$&1.84&1.09&0.937\\
35-40&(8.90$\pm$0.49$\pm$2.49)$\times10^{-1}$&$1.16\pm0.08$&$1.05\pm0.34$&1.16&7.38$\times10^{-1}$&0.936\\
40-50&(4.87$\pm$0.25$\pm$1.26)$\times10^{-1}$&$(5.18\pm0.54)\times10^{-1}$&$(4.89\pm1.67)\times10^{-1}$&6.04$\times10^{-1}$&3.44$\times10^{-1}$&0.915\\
50-70&(1.60$\pm$0.09$\pm$0.40)$\times10^{-1}$&$(1.53\pm0.22)\times10^{-1}$&$(1.60\pm0.51)\times10^{-1}$&2.08$\times10^{-1}$&1.02$\times10^{-1}$&0.966\\
70-90&(5.17$\pm$0.51$\pm$1.41)$\times10^{-2}$&$(3.59\pm0.70)\times10^{-2}$&$(4.24\pm1.21)\times10^{-2}$&5.83$\times10^{-2}$&2.94$\times10^{-2}$&0.954\\
90-120&(1.79$\pm$0.18$\pm$0.50)$\times10^{-2}$&$(9.45\pm2.35)\times10^{-3}$&$(1.25\pm0.30)\times10^{-2}$&1.79$\times10^{-2}$&8.22$\times10^{-3}$&0.920\\
120-170&(4.49$\pm$0.81$\pm$1.58)$\times10^{-3}$&$(1.98\pm0.59)\times10^{-3}$&$(3.13\pm0.51)\times10^{-3}$&4.19$\times10^{-3}$&1.94$\times10^{-3}$&0.907\\
170-300&(6.39$\pm$2.26$\pm$2.04)$\times10^{-4}$&$(1.90\pm0.67)\times10^{-4}$&$(3.99\pm0.25)\times10^{-4}$&4.30$\times10^{-4}$&2.37$\times10^{-4}$&0.913\\
\hline
\multicolumn{7}{c}{$\gamma+c+X$} \\
\hline
30-35&(1.16$\pm$0.05$\pm$0.20)$\times10$&$(1.74\pm0.10)\times10$&$(1.07\pm0.66)\times10$&1.25$\times10$&8.01&1.28\\
35-40&6.33$\pm$0.33$\pm$1.08&$8.82\pm0.72$&$6.22\pm2.77$&7.23&4.39&1.25\\
40-50&2.92$\pm$0.17$\pm$0.48&$3.67\pm0.36$&$2.65\pm1.67$&3.43&2.01&1.21\\
50-70&(7.62$\pm$0.60$\pm$1.39)$\times10^{-1}$&$(8.54\pm1.03)\times10^{-1}$&$(7.26\pm3.02)\times10^{-1}$&9.79$\times10^{-1}$&5.12$\times10^{-1}$&1.16\\
70-90&(1.67$\pm$0.35$\pm$0.37)$\times10^{-1}$&$(1.62\pm0.25)\times10^{-1}$&$(1.71\pm0.54)\times10^{-1}$&2.28$\times10^{-1}$&1.05$\times10^{-1}$&1.13\\
90-120&(4.37$\pm$1.44$\pm$0.85)$\times10^{-2}$&$(3.51\pm0.65)\times10^{-2}$&$(4.99\pm0.97)\times10^{-2}$&5.90$\times10^{-2}$&2.50$\times10^{-2}$&1.11\\
120-170&(1.32$\pm$0.55$\pm$0.26)$\times10^{-2}$&$(5.44\pm1.37)\times10^{-3}$&$(1.25\pm0.02)\times10^{-2}$&1.20$\times10^{-2}$&4.56$\times10^{-3}$&1.07\\
170-300&(1.51$\pm$1.23$\pm$0.45)$\times10^{-3}$&$(3.86\pm1.16)\times10^{-4}$&$(1.92\pm0.10)\times10^{-3}$&1.12$\times10^{-3}$&4.84$\times10^{-4}$&1.04\\
\hline\hline
\end{tabular}\label{tab:results}

\end{center}
\end{table*}

\begin{table}
\caption{Systematic uncertainty on the integrated cross sections.  Effects listed under ``All Others'' include photon energy scale, jet energy scale, and $b$-tagging efficiency.}
\begin{tabular}{l c c}
\hline 
\hline 
Systematic Effect & Uncertainty & Uncertainty \\
 & $\gamma+b+X$ & $\gamma+c+X$ \\
\hline 
 $M_{\rm SecVtx}$ Template &23.2\% & 12.6\%\\
Event Generator & 9.4\% & 5.8\%\\
ANN Template  & 8.9\% & 4.3\%\\
Luminosity & 6.0\% & 6.0\%\\
All Others & 4.5\% & 8.4\%\\
\hline
Total Systematic & 27.6\%  & 17.8\% \\
\hline	
\hline
\end{tabular}
\label{tab:systematics}
\end{table}

The predictions based on NLO pQCD \cite{Stavreva:2009vi} include direct-photon production subprocesses and subprocesses where the photon is emitted from parton fragmentation, both at ${\cal O}(\alpha\alpha_{s}^2)$. The calculation utilizes CTEQ6.6M parton distribution functions \cite{Pumplin:2002vw}. The scale dependence is evaluated by varying the renormalization, $\mu_{r}$, factorization, $\mu_{f}$, and fragmentation, $\mu_{F}$, scales, assumed to be the same, from the default value $p_{T}^{\gamma}$ to $p_{T}^{\gamma}$/2 and 2$p_{T}^{\gamma}$. 

The predictions based on a $k_{T}$-factorization approach \cite{Lipatov:2012rg} include ${\cal O}(\alpha\alpha_{s}^2)$ off-shell amplitudes of gluon-gluon fusion and quark-(anti)quark interaction subprocesses, and the $k_T$-dependent (i.e., unintegrated) parton distributions, where $k_T$ denotes the transverse momentum of the parton. The non-vanishing transverse momentum of the colliding partons leads to a broadening of the photon transverse-momentum distribution. The scale dependence is evaluated in the same way as the NLO calculations.

Both the NLO and $k_{T}$-factorization predictions are parton-level calculations without modeling of underlying-event energy. We correct those two predictions for the non-perturbative effects of parton-to-hadron fragmentation and for underlying-event energy, by multiplying with a correction factor derived from a sample simulated with {\sc sherpa}. The correction factors are shown in Table~\ref{tab:results}.

The predictions of {\sc sherpa} \cite{Gleisberg:2008ta} include all the tree-level matrix-element diagrams  with one photon and up to three jets, with at least one $b$ jet or $c$ jet in the explored kinematic region. This calculation features a parton-jet matching procedure in order to avoid an overlap between the phase-space descriptions given by the fixed-order matrix-element subprocesses and the showering and hadronization in the multi-jets simulation.

The predictions of {\sc pythia} \cite{Sjostrand:2006za} include the $2\rightarrow2$ matrix-element subprocesses $gb\rightarrow\gamma b$ and $q\bar{q}\rightarrow\gamma g$ with $g\rightarrow b\bar{b}$ and $g\rightarrow c\bar{c}$ splittings in the parton shower. In the ratio plots, we multiply the {\sc pythia} calculations by an empirical factor of 1.4 to improve the agreement of the normalization. Previous studies \cite{Abbott:1999se} showed that the contribution of gluon splitting to heavy flavor has to be approximately doubled over expectations from the leading-order {\sc pythia} generator to reproduce the data. Hence, we also show predictions that include a double gluon-splitting rate to heavy flavors.



The NLO pQCD predictions agree with data at low $E_{T}^{\gamma}$ but fail to describe data for $E_{T}^{\gamma} > 70$ GeV for the bottom-jet cross section. The same trend is observed in the charm-jet cross section even though the experimental uncertainty is larger. For large $E_{T}^{\gamma}$, the dominant production process yielding a photon and a heavy quark involves a final-state gluon splitting into a heavy-flavor pair. This process is present only at leading order in the NLO calculation. The {\sc sherpa} prediction allows up to three partons in the final state, through the inclusion of additional tree-level amplitudes. The additional amplitudes also serve as a source of heavy-flavor pairs (through gluon splitting), which is important for the high $E_{T}^{\gamma}$ range. The $k_{T}$-factorization and {\sc sherpa} predictions are in reasonable agreement with the measured cross sections. The {\sc pythia} predictions disagree with the data both in rate and in shape. Scaling the {\sc pythia} prediction and doubling the rate for $g\rightarrow b\bar{b}$ or $g\rightarrow c\bar{c}$ leads to an improved agreement with the data.

In conclusion, we measure the differential cross sections for inclusive production of a photon in association with a heavy flavor quark for $E_{T}^{\gamma}$ between 30 and 300 GeV using the full CDF Run II data set and compare the results with four theoretical predictions.
Most of the models have difficulties in describing the shape of the $E_{T}^{\gamma}$ distribution. The results indicate that an improved understanding of gluon-splitting rates to heavy flavors is important for the NLO pQCD calculations and the {\sc pythia} generator to model data. The results are in agreement with the previous CDF \cite{Aaltonen:2009wc} and D0 \cite{Abazov:2012ea} measurements in the kinematic regions explored. 
These results can be used to improve the background modeling in the searches for new physics in channels involving the production of photons in association with heavy-flavor quarks and to test the models that contain intrinsic heavy quarks.

We are grateful to A.~V.~Lipatov, M.~A.~Malyshev, N.~P.~Zotov, F.~Siegert, T.~P.~Stavreva, and J.~F.~Owens for providing theoretical predictions and for many useful discussions. We thank the Fermilab staff and the technical staffs of the
participating institutions for their vital contributions. This work
was supported by the U.S. Department of Energy and National Science
Foundation; the Italian Istituto Nazionale di Fisica Nucleare; the
Ministry of Education, Culture, Sports, Science and Technology of
Japan; the Natural Sciences and Engineering Research Council of
Canada; the National Science Council of the Republic of China; the
Swiss National Science Foundation; the A.P. Sloan Foundation; the
Bundesministerium f\"ur Bildung und Forschung, Germany; the Korean
World Class University Program, the National Research Foundation of
Korea; the Science and Technology Facilities Council and the Royal
Society, UK; the Russian Foundation for Basic Research; the Ministerio
de Ciencia e Innovaci\'{o}n, and Programa Consolider-Ingenio 2010,
Spain; the Slovak R\&D Agency; the Academy of Finland; the Australian
Research Council (ARC); and the EU community Marie Curie Fellowship
contract 302103.

\end{document}

%% file: Author_List.tex
\affiliation{Institute of Physics, Academia Sinica, Taipei, Taiwan 11529, Republic of China}
\affiliation{Argonne National Laboratory, Argonne, Illinois 60439, USA}
\affiliation{University of Athens, 157 71 Athens, Greece}
\affiliation{Institut de Fisica d'Altes Energies, ICREA, Universitat Autonoma de Barcelona, E-08193, Bellaterra (Barcelona), Spain}
\affiliation{Baylor University, Waco, Texas 76798, USA}
\affiliation{Istituto Nazionale di Fisica Nucleare Bologna, $^{ee}$University of Bologna, I-40127 Bologna, Italy}
\affiliation{University of California, Davis, Davis, California 95616, USA}
\affiliation{University of California, Los Angeles, Los Angeles, California 90024, USA}
\affiliation{Instituto de Fisica de Cantabria, CSIC-University of Cantabria, 39005 Santander, Spain}
\affiliation{Carnegie Mellon University, Pittsburgh, Pennsylvania 15213, USA}
\affiliation{Enrico Fermi Institute, University of Chicago, Chicago, Illinois 60637, USA}
\affiliation{Comenius University, 842 48 Bratislava, Slovakia; Institute of Experimental Physics, 040 01 Kosice, Slovakia}
\affiliation{Joint Institute for Nuclear Research, RU-141980 Dubna, Russia}
\affiliation{Duke University, Durham, North Carolina 27708, USA}
\affiliation{Fermi National Accelerator Laboratory, Batavia, Illinois 60510, USA}
\affiliation{University of Florida, Gainesville, Florida 32611, USA}
\affiliation{Laboratori Nazionali di Frascati, Istituto Nazionale di Fisica Nucleare, I-00044 Frascati, Italy}
\affiliation{University of Geneva, CH-1211 Geneva 4, Switzerland}
\affiliation{Glasgow University, Glasgow G12 8QQ, United Kingdom}
\affiliation{Harvard University, Cambridge, Massachusetts 02138, USA}
\affiliation{Division of High Energy Physics, Department of Physics, University of Helsinki and Helsinki Institute of Physics, FIN-00014, Helsinki, Finland}
\affiliation{University of Illinois, Urbana, Illinois 61801, USA}
\affiliation{The Johns Hopkins University, Baltimore, Maryland 21218, USA}
\affiliation{Institut f\"{u}r Experimentelle Kernphysik, Karlsruhe Institute of Technology, D-76131 Karlsruhe, Germany}
\affiliation{Center for High Energy Physics: Kyungpook National University, Daegu 702-701, Korea; Seoul National University, Seoul 151-742, Korea; Sungkyunkwan University, Suwon 440-746, Korea; Korea Institute of Science and Technology Information, Daejeon 305-806, Korea; Chonnam National University, Gwangju 500-757, Korea; Chonbuk National University, Jeonju 561-756, Korea; Ewha Womans University, Seoul, 120-750, Korea}
\affiliation{Ernest Orlando Lawrence Berkeley National Laboratory, Berkeley, California 94720, USA}
\affiliation{University of Liverpool, Liverpool L69 7ZE, United Kingdom}
\affiliation{University College London, London WC1E 6BT, United Kingdom}
\affiliation{Centro de Investigaciones Energeticas Medioambientales y Tecnologicas, E-28040 Madrid, Spain}
\affiliation{Massachusetts Institute of Technology, Cambridge, Massachusetts 02139, USA}
\affiliation{Institute of Particle Physics: McGill University, Montr\'{e}al, Qu\'{e}bec H3A~2T8, Canada; Simon Fraser University, Burnaby, British Columbia V5A~1S6, Canada; University of Toronto, Toronto, Ontario M5S~1A7, Canada; and TRIUMF, Vancouver, British Columbia V6T~2A3, Canada}
\affiliation{University of Michigan, Ann Arbor, Michigan 48109, USA}
\affiliation{Michigan State University, East Lansing, Michigan 48824, USA}
\affiliation{Institution for Theoretical and Experimental Physics, ITEP, Moscow 117259, Russia}
\affiliation{University of New Mexico, Albuquerque, New Mexico 87131, USA}
\affiliation{The Ohio State University, Columbus, Ohio 43210, USA}
\affiliation{Okayama University, Okayama 700-8530, Japan}
\affiliation{Osaka City University, Osaka 588, Japan}
\affiliation{University of Oxford, Oxford OX1 3RH, United Kingdom}
\affiliation{Istituto Nazionale di Fisica Nucleare, Sezione di Padova-Trento, $^{ff}$University of Padova, I-35131 Padova, Italy}
\affiliation{University of Pennsylvania, Philadelphia, Pennsylvania 19104, USA}
\affiliation{Istituto Nazionale di Fisica Nucleare Pisa, $^{gg}$University of Pisa, $^{hh}$University of Siena and $^{ii}$Scuola Normale Superiore, I-56127 Pisa, Italy, $^{mm}$INFN Pavia and University of Pavia, I-27100 Pavia, Italy}
\affiliation{University of Pittsburgh, Pittsburgh, Pennsylvania 15260, USA}
\affiliation{Purdue University, West Lafayette, Indiana 47907, USA}
\affiliation{University of Rochester, Rochester, New York 14627, USA}
\affiliation{The Rockefeller University, New York, New York 10065, USA}
\affiliation{Istituto Nazionale di Fisica Nucleare, Sezione di Roma 1, $^{jj}$Sapienza Universit\`{a} di Roma, I-00185 Roma, Italy}
\affiliation{Mitchell Institute for Fundamental Physics and Astronomy, Texas A\&M University, College Station, Texas 77843, USA}
\affiliation{Istituto Nazionale di Fisica Nucleare Trieste/Udine; $^{nn}$University of Trieste, I-34127 Trieste, Italy; $^{kk}$University of Udine, I-33100 Udine, Italy}
\affiliation{University of Tsukuba, Tsukuba, Ibaraki 305, Japan}
\affiliation{Tufts University, Medford, Massachusetts 02155, USA}
\affiliation{University of Virginia, Charlottesville, Virginia 22906, USA}
\affiliation{Waseda University, Tokyo 169, Japan}
\affiliation{Wayne State University, Detroit, Michigan 48201, USA}
\affiliation{University of Wisconsin, Madison, Wisconsin 53706, USA}
\affiliation{Yale University, New Haven, Connecticut 06520, USA}

\author{T.~Aaltonen}
\affiliation{Division of High Energy Physics, Department of Physics, University of Helsinki and Helsinki Institute of Physics, FIN-00014, Helsinki, Finland}
\author{S.~Amerio}
\affiliation{Istituto Nazionale di Fisica Nucleare, Sezione di Padova-Trento, $^{ff}$University of Padova, I-35131 Padova, Italy}
\author{D.~Amidei}
\affiliation{University of Michigan, Ann Arbor, Michigan 48109, USA}
\author{A.~Anastassov$^x$}
\affiliation{Fermi National Accelerator Laboratory, Batavia, Illinois 60510, USA}
\author{A.~Annovi}
\affiliation{Laboratori Nazionali di Frascati, Istituto Nazionale di Fisica Nucleare, I-00044 Frascati, Italy}
\author{J.~Antos}
\affiliation{Comenius University, 842 48 Bratislava, Slovakia; Institute of Experimental Physics, 040 01 Kosice, Slovakia}
\author{G.~Apollinari}
\affiliation{Fermi National Accelerator Laboratory, Batavia, Illinois 60510, USA}
\author{J.A.~Appel}
\affiliation{Fermi National Accelerator Laboratory, Batavia, Illinois 60510, USA}
\author{T.~Arisawa}
\affiliation{Waseda University, Tokyo 169, Japan}
\author{A.~Artikov}
\affiliation{Joint Institute for Nuclear Research, RU-141980 Dubna, Russia}
% ...

\collaboration{CDF Collaboration}\altaffiliation[With visitors from]{
$^a$University of British Columbia, Vancouver, BC V6T 1Z1, Canada,
$^b$Istituto Nazionale di Fisica Nucleare, Sezione di Cagliari, 09042 Monserrato (Cagliari), Italy,
$^c$University of California Irvine, Irvine, CA 92697, USA,
$^e$Institute of Physics, Academy of Sciences of the Czech Republic, 182~21, Czech Republic,
$^f$CERN, CH-1211 Geneva, Switzerland,
$^g$Cornell University, Ithaca, NY 14853, USA,
$^{dd}$The University of Jordan, Amman 11942, Jordan,
$^h$University of Cyprus, Nicosia CY-1678, Cyprus,
$^i$Office of Science, U.S. Department of Energy, Washington, DC 20585, USA,
$^j$University College Dublin, Dublin 4, Ireland,
$^k$ETH, 8092 Z\"{u}rich, Switzerland,
$^l$University of Fukui, Fukui City, Fukui Prefecture, Japan 910-0017,
$^m$Universidad Iberoamericana, Lomas de Santa Fe, M\'{e}xico, C.P. 01219, Distrito Federal,
$^n$University of Iowa, Iowa City, IA 52242, USA,
$^o$Kinki University, Higashi-Osaka City, Japan 577-8502,
$^p$Kansas State University, Manhattan, KS 66506, USA,
$^q$Brookhaven National Laboratory, Upton, NY 11973, USA,
$^r$University of Manchester, Manchester M13 9PL, United Kingdom,
$^s$Queen Mary, University of London, London, E1 4NS, United Kingdom,
$^t$University of Melbourne, Victoria 3010, Australia,
$^u$Muons, Inc., Batavia, IL 60510, USA,
$^v$Nagasaki Institute of Applied Science, Nagasaki 851-0193, Japan,
$^w$National Research Nuclear University, Moscow 115409, Russia,
$^x$Northwestern University, Evanston, IL 60208, USA,
$^y$University of Notre Dame, Notre Dame, IN 46556, USA,
$^z$Universidad de Oviedo, E-33007 Oviedo, Spain,
$^{aa}$CNRS-IN2P3, Paris, F-75205 France,
$^{cc}$Universidad Tecnica Federico Santa Maria, 110v Valparaiso, Chile,
$^{ll}$Universite catholique de Louvain, 1348 Louvain-La-Neuve, Belgium,
$^{oo}$University of Z\"{u}rich, 8006 Z\"{u}rich, Switzerland,
$^{pp}$Massachusetts General Hospital and Harvard Medical School, Boston, MA 02114 USA,
$^{qq}$Hampton University, Hampton, VA 23668, USA,
$^{rr}$Los Alamos National Laboratory, Los Alamos, NM 87544, USA
}
\noaffiliation

%% file: cdfgambc.bbl
\begin{thebibliography}{99}
\bibitem{co}
 A cylindrical coordinate system, $(r,\phi,z)$ is used with origin at the geometric center of the detector; $r$ is the radius from the nominal beam line, $\phi$ is the azimuthal angle, and $+z$ points along the direction of the proton beam. The polar angle $\theta$ with respect to the proton beam defines the pseudorapidity $\eta$, which is given by $\eta = - \ln[\tan(\theta/2)]$. Transverse energy and transverse momentum are defined as $E_{T}=E\sin(\theta)$ and $p_{T}=p\sin(\theta)$, respectively.
\end{thebibliography}
